\documentclass[epsf]{mn2e}

\def\mmm{$(m-M)_0$}
\def\ebv{$E(B-V)$~}

\def\gsim{\;\lower.6ex\hbox{$\sim$}\kern-7.75pt\raise.65ex\hbox{$>$}\;}
\def\lsim{\;\lower.6ex\hbox{$\sim$}\kern-7.75pt\raise.65ex\hbox{$<$}\;}

\title[NGC 6939]{UBVI photometry of the intermediate age open cluster NGC 6939}

\author[Andreuzzi et al.]{Gloria Andreuzzi$^{1,2}$, Angela Bragaglia$^3$, 
                          Monica Tosi$^3$, Gianni Marconi$^{1,4}$\\
$^1$ INAF--Osservatorio Astronomico di Roma, Via dell'Osservatorio 5, 
      I-00040 Monte Porzio, Italy, email gloria@mporzio.astro.it\\
$^2$ Telescopio Nazionale Galileo, 38700 Santa Cruz de La Palma, Spain \\
$^3$ INAF--Osservatorio Astronomico di Bologna, Via Ranzani 1, I-40127 Bologna,
      Italy,
      e-mail angela@bo.astro.it, tosi@bo.astro.it \\
$^4$  ESO, Alonso de Cordova 3107, Vitacura, Santiago, Chile,
      e-mail gmarconi@eso.org
}

\date{}

\begin{document}
\maketitle

\begin{abstract}
We present CCD $UBVI$ photometry of the nearby, intermediate age open cluster 
NGC 6939.
Using the synthetic Colour - Magnitude Diagrams technique we estimate the
following parameters: age between 1.3 and 1.0 Gyr (depending on whether or not 
overshooting is considered), reddening 0.34 $\leq$ \ebv $\leq$ 0.38 and 
distance modulus 11.3 $\leq$ \mmm $\leq$ 11.4.

\end{abstract}


\section{Introduction}

A vast amount of  information on the formation and evolution of our Galaxy
can be gathered from the study of open clusters (OCs).
In particular, old OCs may be useful to derive information not only
on the present day situation, but also on the time evolution of the disc,
since their ages cover the whole existence of the disc, reaching to about
10 Gyr (Friel 1995). 
In order to fully exploit the information provided by OCs we
must derive for them in a homogeneous way: accurate absolute ages (and a
consistent age ranking), distances, reddenings, and metal content. With this
aim we are studying a number of old OCs (see Bragaglia \& Tosi 2003 and
Sandrelli et al. 1999  for references), and we now add another cluster to our 
sample. 

Given its relative proximity, NGC 6939 (C2030+604) has been the target of
several studies in the past: the first bibliographic entry is 80 years ago
(Kustner 1923) but, surprisingly, the first CCD data appeared only in 2002
(Rosvick \& Balam 2002, hereafter RB02). As usual, the cluster parameters found
in literature do not agree with each other, and we present in this paper new
and improved determinations for this intermediate age open cluster located at
RA(2000) = 20:31:32, DEC(2000) = +60:39:00, or l = 95.88,  b = 12.30. 

Photometry has been previously presented by several authors, but old
photographic photometry only reached about one magnitude below the main
sequence Turn-Off.  
Mermilliod, Huestamendia, \& del Rio (1994) took UBV photoelectric photometry
of 37 member stars  all in the red clump phase, with the intent of
discriminating between different evolutionary models (with or without
overshooting) by comparison with theoretical  isochrones.  It turned out that
most of the bright stars lying in the cluster direction  are indeed cluster
members (their  Table 1 lists only 4 non-members among 45 objects). 
The recent work by Rosvick \& Balam (2002) has presented the first deep BVI CCD
data.  They used the 1.85m Dominion Astrophysical Observatory telescope,
covering the central part of NGC 6939 (more or less our field A, see later).
Their CMDs show considerable scatter, which they do not attribute to
contamination from field stars (even if they could not prune their diagrams on
the main sequence, since no proper motion survey on this cluster has reached so
deep), but to differential reddening [\ebv = 0.29 to 0.41], which also
influences the distance derivation [$(m-M)_V$ = 12.21 to 12.39]. Using the
Girardi et al. (2000) solar metallicity isochrones, they obtain a cluster age
of 1.6 $\pm$ 0.3 Gyr

Canterna et al. (1986)  give a metal abundance of [Fe/H] = --0.1 based on 
Washington photometry, a value confirmed by the revision of this method done by
Geisler, Clari\'a \& Minniti (1991), who give  --0.13.
Thogersen, Friel \& Fallon (1993) determine a cluster mean radial velocity of
$-42 \pm 10$ kms$^{-1}$, and [Fe/H]=--0.14$\pm$0.13, from low resolution
spectra of four cluster giants. The metallicity has recently been revised to
[Fe/H]=--0.19$\pm$0.09 by Friel et al. (2002).
Other measures of radial velocities were given by Geisler (1988), who found
$-24.1 \pm 3.1$ based on medium resolution spectroscopy of 8 stars; by  Milone
(1994) who, on the basis of higher resolution spectra of 26 giants, found a
mean cluster radial velocity of $-19.89 \pm 0.19$ kms$^{-1}$; and by  Glushkova
\& Rastorguev (1991), who derived $-19.3$ as average for 29 cluster members.
These last also determined proper motions (hence membership) for 136 stars in
the cluster vicinity. 
Zhao et al. (1985) determined membership probabilities based on proper motion
for more than 200 stars in the cluster field, and we used their information
to confirm the turn-off and red clump positions in our data.

Dutra and Bica (2000) cite a  reddening  \ebv=0.39 based on the Schlegel,
Finkbeiner \& Davis (1998) maps.

At least two groups used existing literature data to re-derive relevant
properties.
Twarog, Ashman \& Anthony-Twarog (1997), in their work trying to derive on
homogeneous grounds the properties of Galactic open clusters, found
\ebv = 0.45 to 0.49, [Fe/H] = 0.03, distance modulus, 
$(m-M)_V$ = 12.45.\footnote{The value listed in their Table 2
  is 13.05, but Bruce Twarog kindly informed us that their correct modulus is
  12.45, because, due to a typo, 13.05 is actually not the modulus but the mean
  visual magnitude of the red clump.}
Carraro, Ng \& Portinari (1998) derived, using the synthetic CMD techniqe with
the Padova tracks and the Thogersen et al. (1993) metallicity, an age of
1.4 Gyr, and  R$_{\rm GC}$ = 9.7 kpc.  

Finally, Robb \& Cardinal (1998) tried to detect variables in this cluster and
actually found six, of which at least two are eclipsing variables.

We describe our data in Section 2, and the resulting colour - magnitude
diagrams in Section 3. Section 4 is devoted to the derivation of the cluster
parameters, while summary and conclusions are presented in Section 5.

\begin{figure}
\vspace{14cm}
\includegraphics{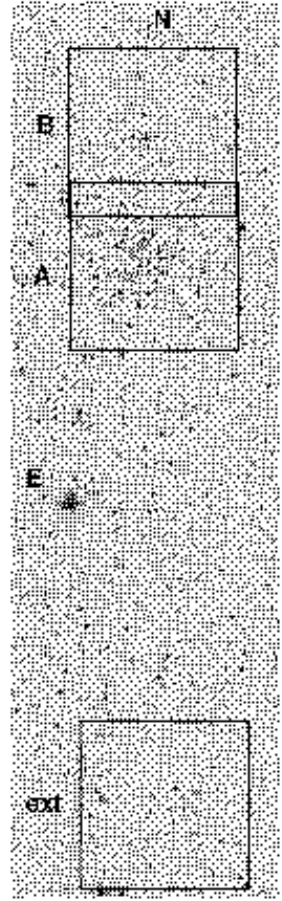}
\caption{Position of our three fields (A, B and external), each one of 9.4
$\times$9.4 arcmin; the map is 15 x 50 arcmin, and is oriented with north up and East left.} 
\label{fig-map}
\end{figure}

\begin{figure}
\vspace{12.5cm}
\includegraphics{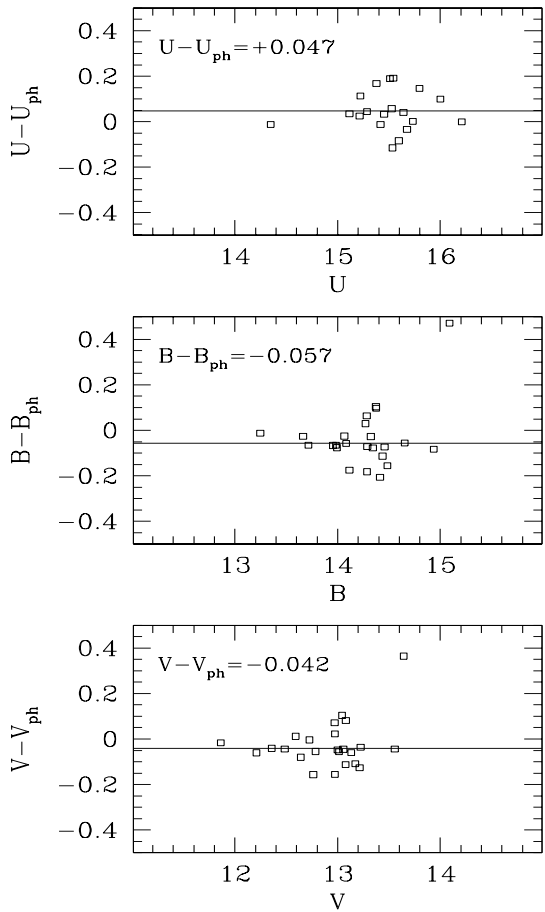}
\caption{Comparison of our U,B,V photometry with the photoelectric
values given in Mermilliod et al. (1994).} 
\label{fig-photo}
\end{figure}

\begin{figure}
\vspace{12cm}
\includegraphics{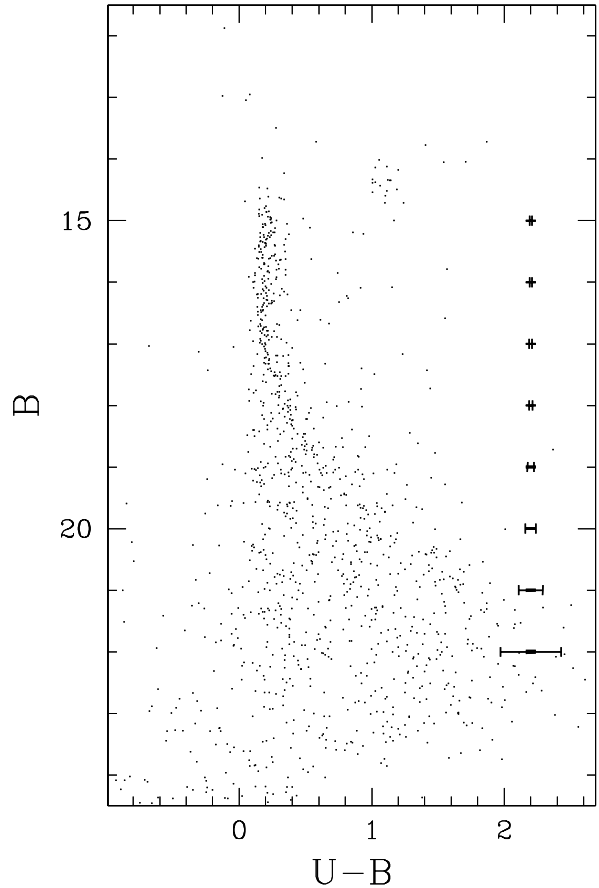}
\caption{B, U-B CMD for NGC 6939.The mean errors per interval of
         magnitude B are also plotted.} 
\label{fig-cmd1}
\end{figure}

\section{Observations and data reduction}

Our data were acquired at the Telescopio Nazionale Galileo, on the Canary
Islands, using  DOLORES (Device Optimized for the LOw RESolution), a focal
reducer capable of imaging and low resolution spectroscopy, on UT 2000
November 24, 25, and 26. A few additional images of very short exposure
time were kindly obtained by M. Bellazzini on UT 2001 August 17 and 18.
In both cases DOLORES
mounted a 2k Loral thinned and back-illuminated CCD, with scale of 0.275
arcsec/pix, and a field of view 9.4 $\times$ 9.4 arcmin$^2$. 
We observed three fields (see Fig.~\ref{fig-map}), one
centered on the  cluster (field A), one slightly North of it (field B), and one
about 30 arcmin away, to be used for field stars decontamination.
This distance was judged reasonably safe, since the cluster diameter
was given as 10 arcmin in the BDA compilation (Mermilliod et al. 1995).
 
In particular we used the region in common between the fields A and B to obtain
a homogeneous calibration for the whole  cluster sample.
For each field, Table 1 lists  the date of the observation
together with the filters used and the corresponding ranges of 
exposure time (in seconds).
Of the three nights, the  first was photometric; seeing conditions were not
optimal for the site, but were reasonable during the observations of fields A
and B (1.1 $\arcsec$ to 1.8 $\arcsec$, with a mean value around 1.4 $\arcsec$),
and degraded (2$\arcsec$ to 3$\arcsec$) only for the external field, the less crowded
one.

\begin{table*}
\begin{center}
\caption{Log of our observations, with coordinates, dates and range in exposure
times (in seconds) for each filter.}
\vspace{5mm}
\begin{tabular}{lclllll}
\hline\hline
Field & Equat. coords (2000) & UT dates & t$_U$ & t$_B$ & t$_V$ & t$_I$\\
\hline
Field A  &20:31:31 +60:29:21 &2000 Nov 24, 25 &60-1200 & 600-20 & 300-60 & 300-10\\
         &		     &2001 Aug 17     &        & 5-2 & 10-2 & 5-2  \\
Field B  &20:31:30 +60:46:51 &2000 Nov 26     &        &600-40 & 300-20 & 300-20\\
         &		     &2001 Aug 18     &        & 5-2 & 5-2 & 5-2\\
external &20:31:32 +60:09:23 &2000 Nov 25     &        & 600-20 & 300-10 & 300-20\\
\hline
\end{tabular}
\end{center}
\end{table*}

\subsection{Data reduction}

Corrections to the raw data for bias, dark and flat-fielding
were performed using the standard IRAF routines.
Subsequent data reduction and analysis was done following the same procedure
for the three data-sets and using the DAOPHOT-ALLFRAME packages
(Version 3, Stetson 1997) with a quadratically varying point spread
function (PSF).
In particular the deep I images of each field were used to find the objects
4 $\sigma$ above the background; the identified
candidates were measured on each of the individual I, V, B and U frames of each
field. 
Two final catalogs 
including all the objects in all the filters have been created: one
comprises only stars in fields A and/or B, the other only stars in the
external field.
As we observed with very different exposure times (see Table 1), 
we decided to include in our final catalog also the 
objects for which we had only one measure available (i.e., the brightest stars,
which are saturated in all frames except the shortest ones).
The objects in common between the fields A and B were used to
homogenize the photometry of the cluster (computing the weighted average of
magnitudes in each filter).
The final cluster catalog includes 4547 objects identified in at least 
one filter. 

\subsection{Photometric calibration}

We observed the standard areas PG0231+051 and Rubin 149, plus
two isolated stars, G156--31 and G26--7 (Landolt 1992), several times
during the two nights. The 16 stars retained in our calibrations have colours
$ -1.192 < U-B < 1.342, -0.329 < B-V < 1.993, -0.534 < V-I < 1.951$, 
that generally well match the ones for cluster stars, and require 
some extrapolation only at the red limit, i.e., for the fainter stars. 

The calibration equations were derived assuming the average extinction
coefficients for the site, as given in the site web pages (www.tng.iac.es:
$\kappa_U=0.49, \kappa_B=0.25, \kappa_V=0.15, \kappa_I=0.07$) and are in the
form:
$$  U = u +0.1796 \times (u-b) -0.9120  ~~(rms=0.043) $$
$$  B = b +0.0525 \times (b-v) +1.4187  ~~(rms=0.016) $$
$$  V = v -0.1490 \times (b-v) +1.2389  ~~(rms=0.014) $$
$$  I = i +0.0282 \times (v-i) +0.7854  ~~(rms=0.015) $$
where u, b, v, i, are the aperture corrected instrumental magnitudes, after
further correction for extinction
and for exposure time, and U, B, V, I are the
output magnitudes, calibrated to the Johnson-Cousins standard system. 

We checked our calibration in two ways. First we compared the equations with
the average ones given for the site and instrument, found in the TNG web pages;
they are not strictly identical, but the coefficients have the same signs and 
similar values. Of course when we apply the two different equation sets, we do
not obtain the same magnitudes (differences are of about 0.2 - 0.3 mag for the
average colours of our stars) but we did not expect more than a general
agreement, and we will stick to the equations presented here, derived exactly
for the nights of our observations.

Second, for 24 stars in common, we directly compared our U, B, V  magnitudes
with the photoelectric values given in Mermilliod et al. (1994), and results
are presented in Fig.~\ref{fig-photo}. No trend is present, but only a shift; 
in particular: $U = U_{ph} +0.047$ ($\sigma_U$ = 0.086), $B = B_{ph} -0.057$ 
($\sigma_B$ = 0.081), $V = V_{ph} -0.042$ ($\sigma_V$ = 0.069).
We then decided to correct our magnitudes using these values; nothing could
be done with I, not present in the photoelectric catalogue, but our calibration
seems reasonable anyway (see the comparison with RB02 in the next Section).

Finally, we note that we did not have any problem with shutter timing, 
since they arise only below 0.1 seconds with this instrument, and our 
shortest exposure was 10 times more for the standard stars (and 20 times 
more for NGC 6939).

\section{Data analysis}

\subsection{The colour - magnitude diagram}

\begin{figure*}
\vspace{12.5cm}
\includegraphics{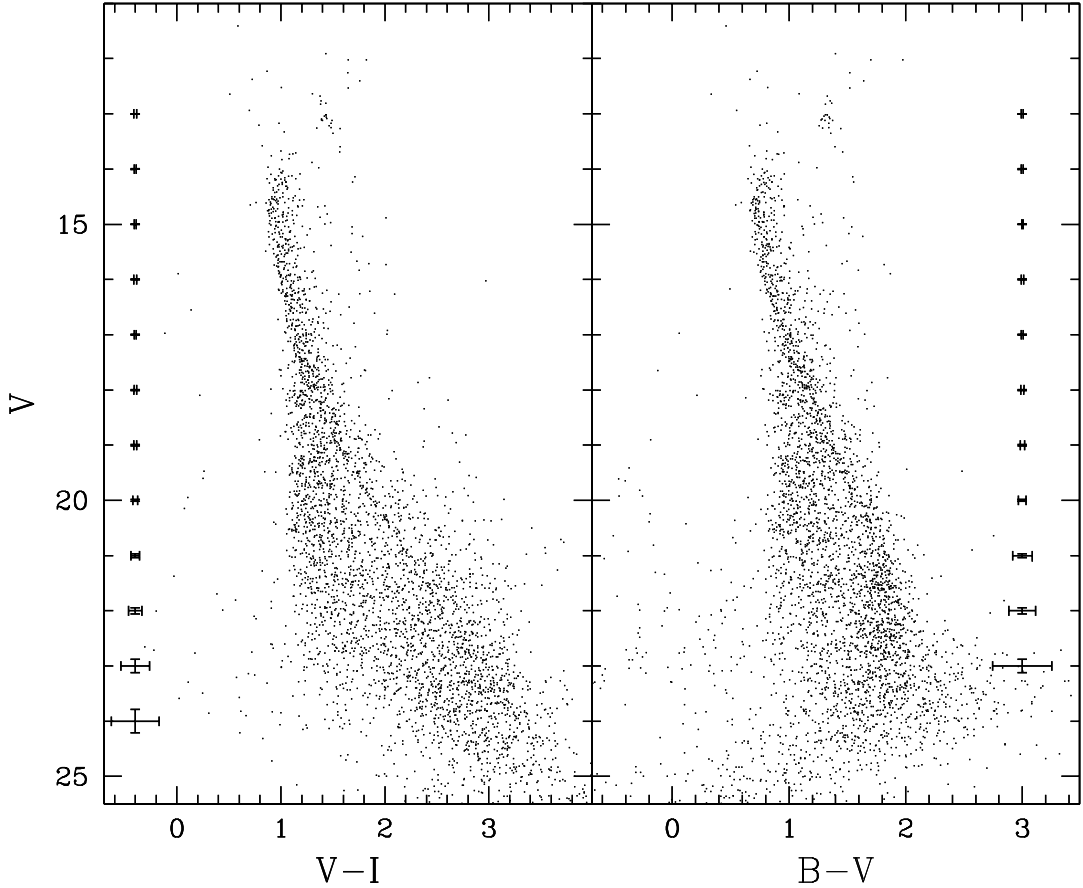}
\caption{Left panel: V, V-I CMD for NGC 6939; Right panel: V, B-V CMD. 
        The mean errors per interval of magnitude V are also plotted. 
	 } 
\label{fig-cmd2}
\end{figure*}

The final, calibrated sample of the cluster consists of: 3796
objects simultaneously identified in the filters V and I, 3804 objects
identified in the filters B and V, and 1276 objects identified in the
filters U and B.
The corresponding CMDs are shown in Figs. \ref{fig-cmd1} and \ref{fig-cmd2},
where the mean errors per interval of magnitude are also plotted.

The CMDs show: i) a very clear MS extending down to V $\sim$ 24,
nearly 5 mag deeper than the only previous CCD photometry (see RB02);
ii) a prominent red giant clump, including $\sim$ 40 stars. 
What we claim is a fairly well 
defined White Dwarf (WD) cooling sequence extending down to V $\sim$ 23.5. 

In principle, the WD cooling sequence may be used to get information 
about a number of astrophysical questions.
Stellar evolution theory predicts that all single stars having a 
MS mass lower than $\simeq$ 8 M$_\odot$ end their lives as WDs. 
Cluster  WD cooling sequences could be thus used to give
an estimate of the initial mass - final mass relationship 
between the stellar progenitor mass from near 8 M$_{\odot}$ to the 
current-day TO mass, and thus about the amount of mass lost from stars during 
their evolution through stellar winds and planetary nebula ejection.
Since the WDs cool at a well-known rate, their cooling sequence may also be used 
to estimate the cluster age. 

However, to use WDs as age estimators one needs 
a) a large sample of WDs to define  the WD luminosity function sufficiently 
well and estimate the turndown of the WD luminosity function with a precision 
equivalent to a Gyr of cooling time; 
b) to separate the cluster WDs from the very high number of blue galaxies 
mixing with WDs at such faint magnitudes.
By looking at the WDs cooling sequence appearing in the NGC\,6939 CMDs it
is evident that, in our case, it is not well enough sampled and well defined down 
to the faint end to use this feature of the CMD as an age estimator.

\begin{figure*}
\vspace{12.5cm}
\includegraphics{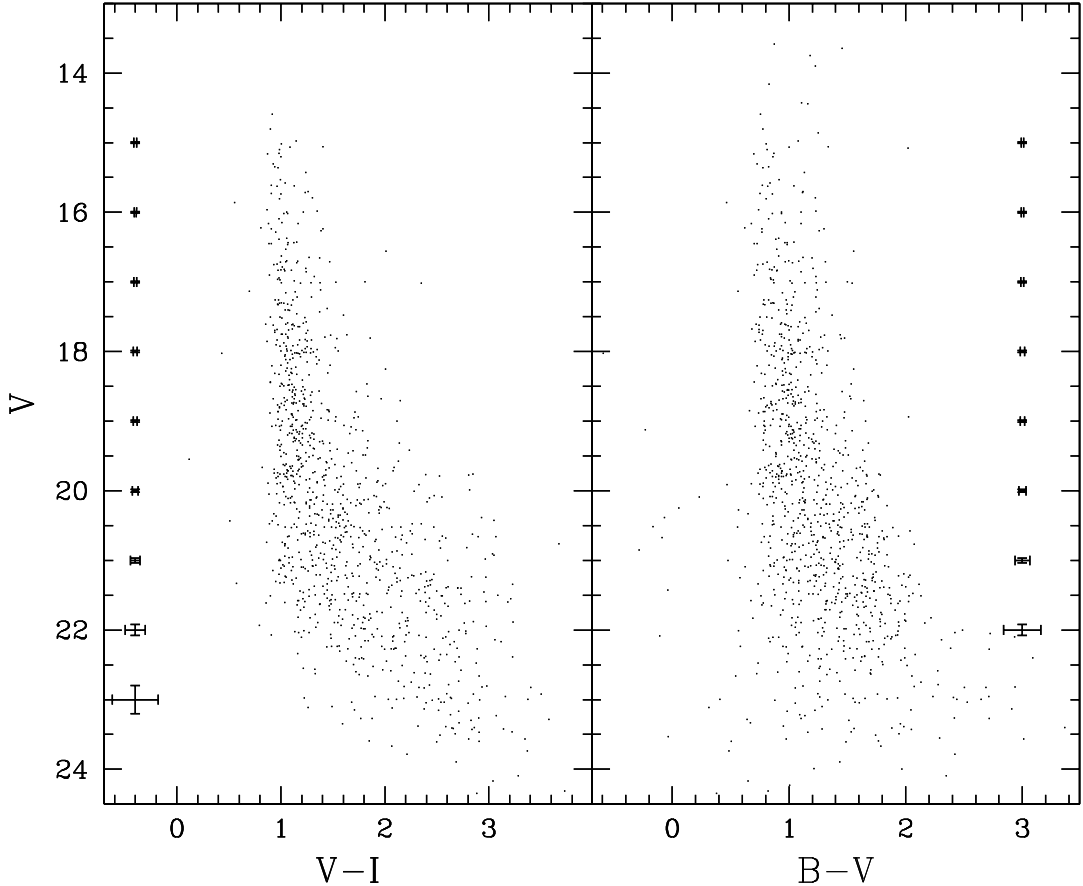}
\caption{Left panel: V, V-I CMD for the external field;
         Right panel: V, B-V CMD for the same field. The mean errors
	per interval of magnitude V are also plotted } 
\label{fig-cmd3}
\end{figure*}

From the CMDs a contamination from field stars is also evident, especially
in the fainter parts. 
In order to account for field stars contamination, we used
data from the field observed far enough from the cluster so
that the contamination by cluster members - if any - is minimal. 
Figure \ref{fig-cmd3} shows the V, V-I and V, B-V CMDs for 1007 
objects simultaneously identified in the filters I and V (left panel) and 
1029 in the filters B and V (right panel) for our external field.

\begin{figure*}
\vspace{9cm}
\includegraphics{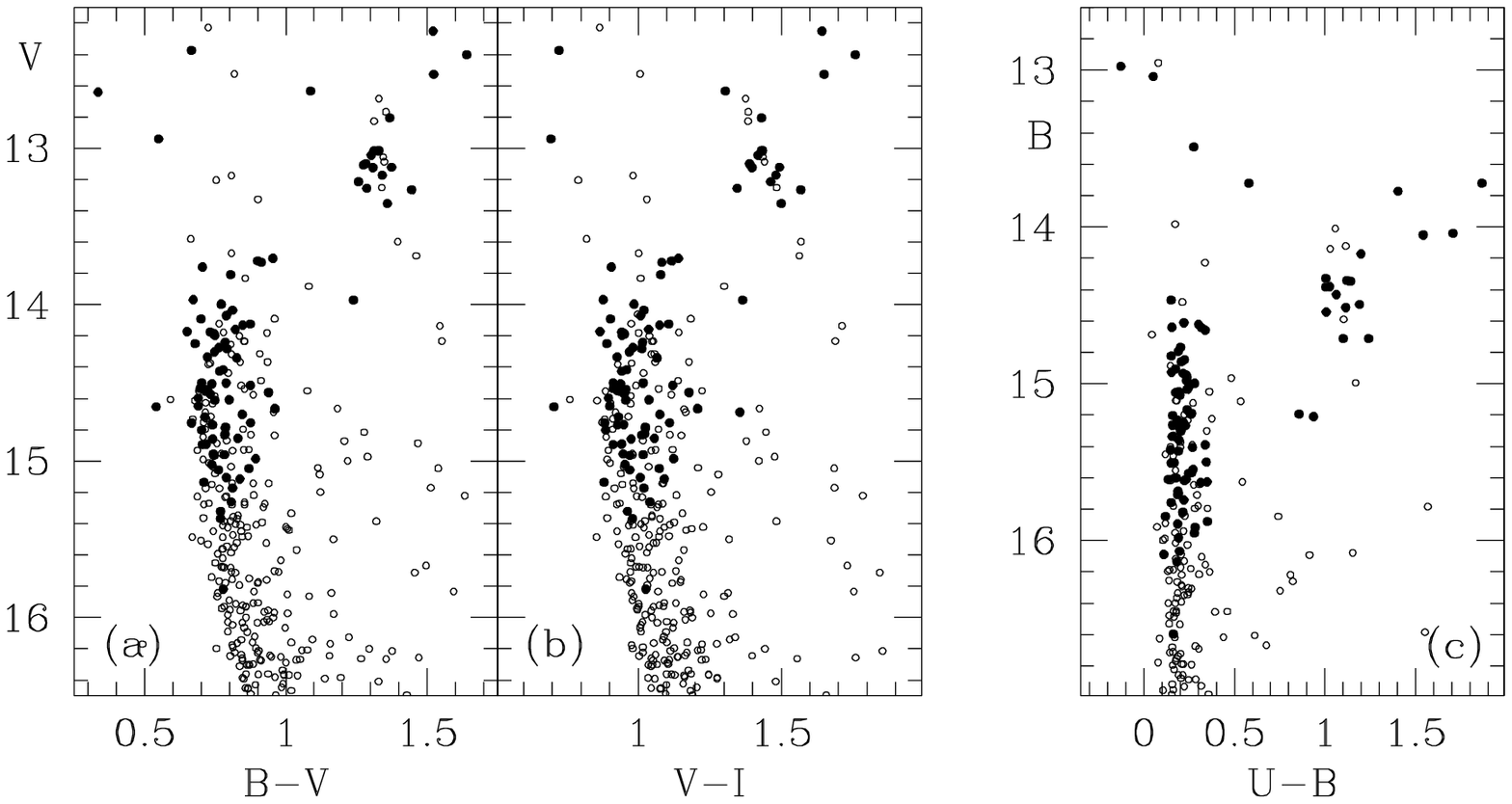}
\caption{Stars in the B-V, V-I and U-B
cluster CMDs (empty points). Filled points represent objects of our
catalog for which the probability to be a member of the cluster, according
to their proper motions, is p $\geq$ 0.9.
} 
\label{fig-mem}
\end{figure*}

Ideally, one would like to eliminate the contamination from the stars of  the
galactic field  using the information contained in the proper motions of the
stars. Unfortunately, no proper motion survey to date reaches the faint
limits required to completely clean our CMD's, but we can at least obtain
information on the bright part and isolate bona fide cluster TO and red clump 
stars.
To do that, we identified 119 objects in common with the catalog of 
Zhao et al. (1985), that gives membership probabilities 
according to proper motion.
The results  are shown in Fig. \ref{fig-mem} where
filled points represent the objects of our catalog for
which the probability to be a member of the cluster is $\geq$ 0.9.

\subsection{Completeness analysis}

Incompleteness factors were determined for each field by carrying
out artificial star tests separately on all
the $I$, $V$, $B$ and $U$ frames, in order to be able to estimate the overall
completeness of our final CMD.

Artificial stars were first added randomly to the $I$, $B$, $V$ and $U$ images
taken as reference in the original photometry. We followed the observed
luminosity function to define the artificial stars magnitudes,  paying
attention to add only a few percent ($\leq$ 10 $\%$) of the total number of
stars actually present in the corresponding magnitude bin to avoid a 
significant enhancement of the image crowding. Then the same stars were added
to the other $I$, $V$, $B$ and $U$ frames in the same positions of the
reference frames, and with the same magnitudes. Finally, all frames  were 
treated with the same analysis adopted for the original ones, and for each
filter and in each magnitude bin a final catalog of artificial stars has been
created.

If a star in the output catalog was found with output values consistent with
the input ones
($\Delta y, \Delta y < 1.5$ pixel, $\delta mag <$ 0.75),
 it was added to the list of recovered objects. 
The ratio of the recovered over added stars,  $\Phi$ = N$_{\rm rec}$/N$_{\rm
add}$ (the completeness factor)  was then derived  after 12 to 15 trials for
each filter and each magnitude bin,  for a total of about 12000 artificial
stars. The estimated completeness factors for the three fields are listed in
Table \ref{tab-compl}.
The quite different completeness fractions found for the central fields
and the external one are due to the seeing: we observed the  external
field in much worse seeing conditions.

\begin{table*}
\caption{Completeness ratios in the four bands for the 2 fields centered
on the cluster and the external one. The first 4 columns give the centres 
of the magnitude bins in each filter; the next columns are
the per cent completeness for field A, for field B and
for the external field, respectively.}
\vspace{5mm}
\begin{tabular}{ccccrrrrrrrrrr}
\hline
\multicolumn{4}{c}{}
&\multicolumn{4}{c}{Field A} &\multicolumn{3}{c}{Field B}
&\multicolumn{3}{c}{External}\\
  U   &  B   & V    & I    &  cU  & cB  &cV   &cI    &  cB  & cV  & cI  &cB &cV &cI \\
\hline
15.01 &16.99 &16.03 &15.71 & 1.00 &1.00 & 1.00&0.98  & 1.00 &1.00 &1.00 &1.00 &1.00 &1.00 \\
15.51 &17.49 &16.53 &16.21 & 0.98 &1.00 & 1.00&0.96  & 1.00 &1.00 &1.00	&1.00 &1.00 &1.00 \\
16.01 &17.99 &17.03 &16.71 & 1.00 &0.90 &0.92 &0.98  & 0.90 &0.90 &1.00	&0.90 &1.00 &0.97 \\
16.51 &18.49 &17.53 &17.21 & 0.98 &0.88 &0.94 &0.99  & 0.90 &0.96 &1.00	&0.80 &1.00 &0.98 \\
17.01 &18.99 &18.03 &17.71 & 1.00 &0.86 &0.96 &0.96  & 0.88 &0.98 &0.99 &0.88 &0.98 &0.92 \\
17.51 &19.49 &18.53 &18.21 & 1.00 &0.86 &0.94 &0.94  & 0.90 &1.00 &0.98 &0.82 &0.97 &0.96 \\
18.01 &19.99 &19.03 &18.71 & 0.90 &0.83 &0.96 &0.95  & 0.90 &0.99 &0.94 &0.80 &0.86 &0.86 \\
18.51 &20.49 &19.53 &19.21 & 0.95 &0.80 &0.94 &0.92  & 0.88 &0.98 &0.98 &0.85 &0.93 &0.96 \\
19.01 &20.99 &20.03 &19.71 & 0.99 &0.84 &0.89 &0.88  & 0.90 &0.97 &0.97 &0.81 &0.90 &0.82 \\
19.51 &21.49 &20.53 &20.21 & 0.97 &0.78 &0.95 &0.86  & 1.00 &0.99 &0.93 &0.77 &0.87 &0.72 \\
20.01 &21.99 &21.03 &20.71 & 0.87 &0.77 &0.95 &0.75  & 0.84 &0.96 &0.94 &0.73 &0.65 &0.51 \\
20.51 &22.49 &21.53 &21.21 & 0.89 &0.83 &0.95 &0.59  & 0.89 &0.93 &0.83 &0.61 &0.58 &0.25 \\
21.01 &22.99 &22.03 &21.71 & 0.66 &0.79 &0.89 &0.30  & 0.85 &0.88 &0.62 &0.44 &0.46 &     \\
21.51 &23.49 &22.53 &22.21 & 0.50 &0.62 &0.78 &      & 0.84 &0.95 &0.22 &0.22 &0.23 &     \\
22.01 &23.99 &23.03 &23.71 & 0.29 &0.43 &0.63 &      & 0.66 &0.80 &     &     &     &     \\
\hline
\end{tabular}
\label{tab-compl}
\end{table*}

\subsection{Comparison with RB02}

\begin{figure}
\vspace{12.5cm}
\includegraphics{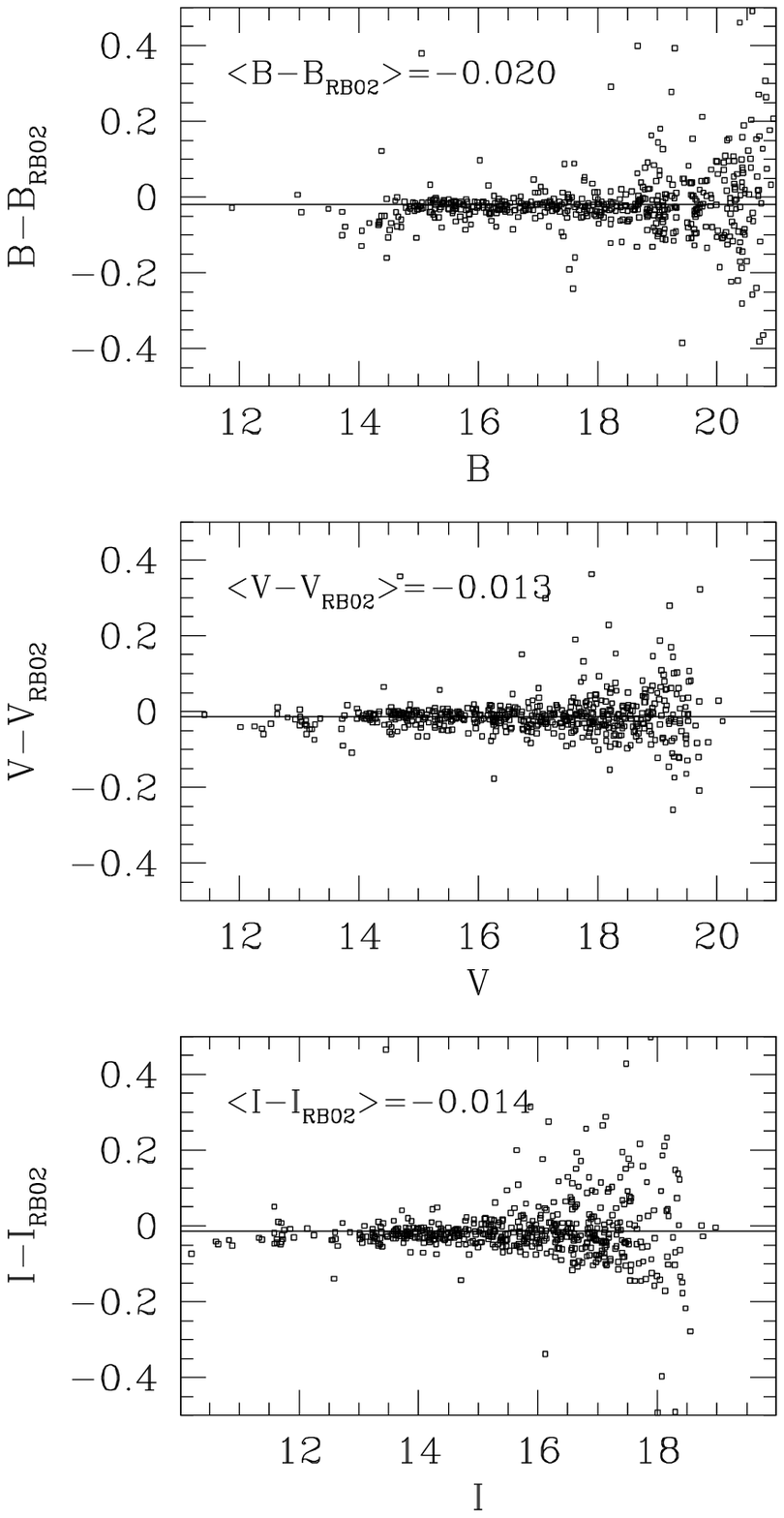}
\caption{Comparison of our B,V,I photometry with the one in RB02.} 
\label{fig-confrb02}
\end{figure}

NGC 6939 has three photographic photometries published, and only one
CCD (RB02). We have compared our B, V, I values with the last one, since
it is the most recent and deepest one.
We have cross-identified stars in the two catalogues using a software
written by P. Montegriffo at the Bologna Observatory, and found 540 objects
in common. The comparison of ours and their magnitudes shows that no 
obvious trend is present over the whole magnitude range.
The magnitude differences between the two catalogues (discernible only at the bright
ends) range between 0.013 and 0.020, which is about the size of our photometric 
errors (see Fig.~\ref{fig-confrb02}).

RB02 find indications of differential reddening for this cluster (see
Introduction), but from our data we cannot confirm it. To look for this effect
we have divided the central field into 9 parts (about 700 $\times$ 700 pixel
each) and plotted the relative  CMDs. Of course the number of stars is quite
small except for the very central part of the cluster, but we find that the TO
position seems to be the same in all cases, both in magnitude and colour. If
differential reddening is present, it is not at the level claimed by RB02.
Furthermore, also the synthetic CMDs do not support this hypothesis (see
Section 4).

\section{Cluster parameters}

Age, reddening and distance of NGC 6939 have been derived applying the
synthetic CMD method (Tosi et al. 1991) to the empirical CMD obtained from the
photometry described above. The best values of the parameters are found by
selecting those providing synthetic CMDs with morphology, number of stars
in the various evolutionary phases and luminosity functions (LF) in best
agreement with the empirical ones.

As already done in the previous papers
of this series (Bragaglia \& Tosi 2003, Di Fabrizio et al. 2001,  Sandrelli
et al. 1999 and references therein), to estimate the error on the
parameters resulting from the intrinsic uncertainties on stellar evolution
theory, the method has been applied adopting various sets of stellar models,
computed by different groups with different assumptions. 
Since the metallicity attributed to NGC 6939 ranges between solar and slightly
sub-solar (see Introduction), we have computed the synthetic CMDs assuming
various metallicities to check which ones provide stellar distributions (i.e.
morphology and number counts) in the CMD in best agreement with the data. 
The adopted sets of stellar tracks are listed in Table 3, where the 
corresponding references are also given, as well as the  model metallicity and
the information on whether or not they take overshooting from convective
zones into account. 

\begin{table}
\begin{center}
\caption{Stellar evolution models adopted for the synthetic CMDs}
\vspace{5mm}
\begin{tabular}{cccl}
\hline\hline
   Set  &metallicity & overshooting & Reference \\
\hline
BBC & 0.02 & yes &Bressan et al. 1993 \\
BBC & 0.008& yes &Fagotto et al. 1994 \\
FRA & 0.02 & no &Dominguez et al. 1999 \\
FRA & 0.01 & no &Dominguez et al. 1999 \\
FST & 0.02 & $\eta$=0.02 &Ventura et al. in prep.\\
FST & 0.02 & $\eta$=0.03 &Ventura et al. in prep.\\
FST & 0.006 & $\eta$=0.02 &Ventura et al. in prep.\\
FST & 0.006 & $\eta$=0.03 &Ventura et al. in prep.\\
\hline
\end{tabular}
\end{center}
The FST models have been kindly provided in advance of publication. 
A description of the corresponding stellar evolution code and assumptions is 
given by Ventura et al. (1998).
\label{tracks}
\end{table}

\begin{figure}
\vspace{15cm}
\includegraphics{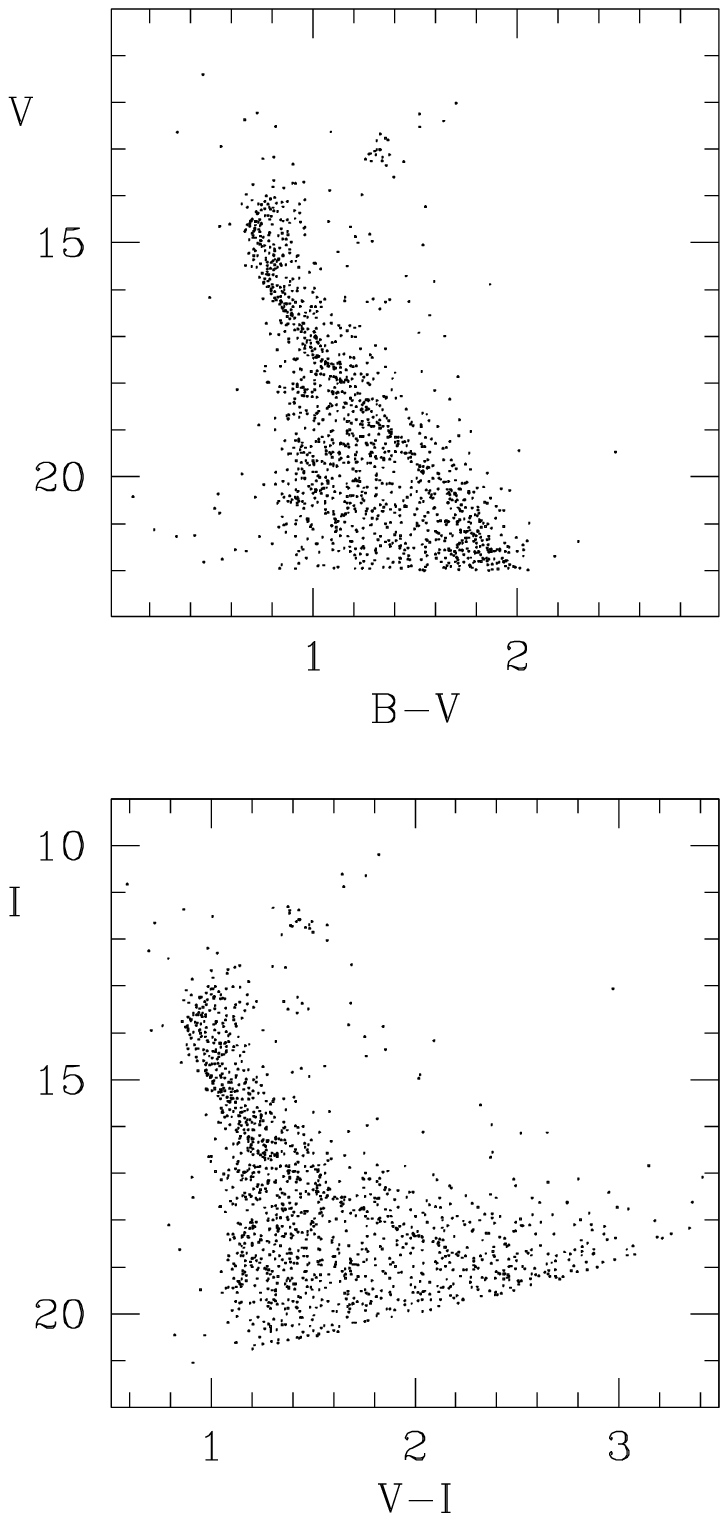}
\caption{CMDs adopted for the comparison with the synthetic ones. The 1439 
plotted stars are located within 6$\arcmin$ from the cluster centre and are 
brighter than V=22 (see text). }
\label{empcmd}
\end{figure}

Having proper motion information (Zhao et al. 1985) on the cluster membership
only for 119 relatively bright stars, to maximize the 
probability of examining only likely members, we have considered only the 
stars located within a 6$\arcmin$ radius from the cluster centre.  Since the
external field, observed for background/foreground decontamination, has a
magnitude limit of V=22, we have also limited the simulations to stars brighter
than this magnitude. The corresponding observational CMDs adopted for the
comparison with the synthetic ones are shown in Fig.~\ref{empcmd}.
Of these 1439 stars, 638 are considered cluster members because they are 
located on the MS, TO, subgiant and clump regions of the CMD.

The synthetic CMDs therefore contain 638 stars down to V=22, extracted with a 
MonteCarlo procedure from the adopted sets of stellar evolution tracks and 
selected according to the B, V and I completeness factors listed in Section 
3.2. To these stars we attributed  the same photometric errors as the actual 
data.  The transformations
from the theoretical luminosity and effective temperature to the
Johnson-Cousins magnitudes and colours have been performed  using Bessel, 
Castelli \& Pletz (1998) conversion tables and assuming $E(V-I)$ = 1.25 \ebv
(Dean et al. 1978) for all the sets of models. Hence the different results
obtained with different stellar models are only attributable to the models
themselves and not to the photometric conversions.
 
In all cases we have found a small internal inconsistency between the B--V and 
V--I colours. When the observed B--V colours are reproduced, the synthetic 
V--I are systematically too blue by about 0.02. This problem is not solved by
adopting different $E(V-I)$/\ebv ratios. The turn-off B--V colours can be 
brought into perfect agreement with the V--I ones if one considers that
reddening is not the same for all stars but depends on their colour (e.g.
Twarog, Ashman \& Anthony-Twarog 1997). However, the dependence suggested by
Fernie (1963) and adopted by Twarog et al. leads to a steepening of the
synthetic main-sequence (MS) that makes the MS shape and the colours of its 
faint stars too blue and even more inconsistent with the data. On the other hand,
it is very likely that the colour offset is simply due to the circumstance that 
the calibration of our I frames is less robust than that in the other bands, 
because there were no photoelectric measurements available in I. In the
selection of the best cluster parameters, we then give higher weight to the 
V, B--V diagrams than to the V, V--I ones, and to the upper MS and clump
regions than to the lower MS (disagreement between this last and the theoretical
predictions is not uncommon, see next Section).

The synthetic CMDs have been computed either assuming that all the cluster
stars are single objects or that a fraction of them are members of binary 
systems with random mass ratio. We find that a binary fraction around 30\% 
well reproduces the observed distribution and spread of the cluster upper main 
sequence. The lower main sequence of the synthetic diagrams comes out
systematically thicker than the observed one, but this inconsistency is
independent of the assumed fraction of binaries and seems rather to suggest that
we have overestimated the photometric errors at faint magnitudes. 
Vice versa, the TO and clump regions of the synthetic CMDs are always 
tighter that the observed ones. This is partially ascribable to field 
contamination and, as far as the clump is concerned, to variable mass 
loss rates affecting the actual stars and not included in the simulation code.

We have checked the suggestion by RB02 of a significant differential reddening.
Synthetic CMDs (with or without binaries) are consistent with the observed MS
width only if $\Delta$\ebv $\leq$0.03. Beyond this difference the MS widens too
much and the TO region splits into separate portions.

\begin{figure*}
\vspace{15cm}
\includegraphics{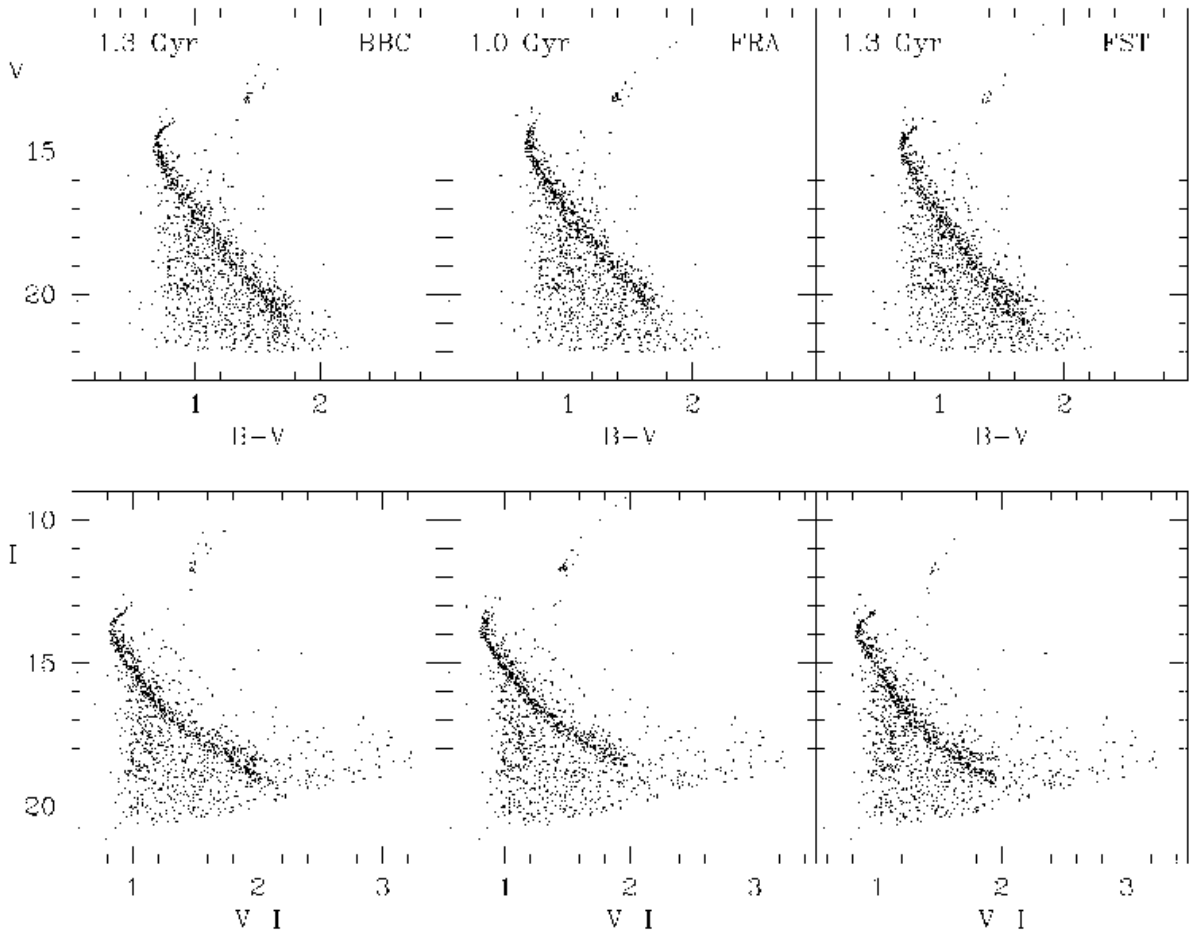}
\caption{Synthetic CMDs in best agreement with the observational ones. 
Left-hand panels: V, B--V (top) and I, V--I (bottom) CMDs from solar 
metallicity BBC models with overshooting; central panels: from solar 
metallicity FRA models without overshooting; right-hand panels: from solar
metallicity FST models with moderate overshooting.  }
\label{simpap}
\end{figure*}

\begin{figure}
\vspace{5cm}
\includegraphics{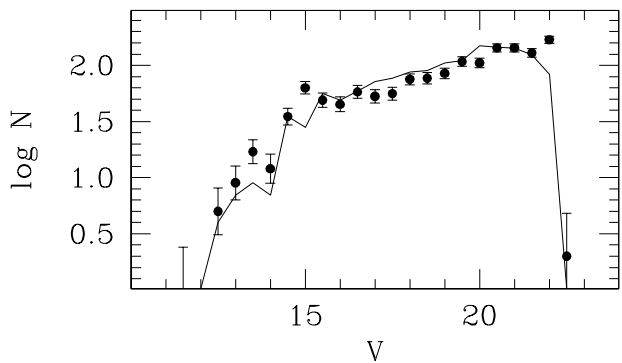}
\caption{V luminosity function of the reference observed stars of
Fig.\ref{empcmd} (dots) and of a typical synthetic case in good agreement with 
the data (line). Errorbars reflect the counting uncertainty as derived from a 
Poisson distribution.}
\label{lf}
\end{figure}

The best synthetic CMDs are chosen on the basis  of the morfology and the number 
counts of the various evolutionary phases ( e.g. MS slope, 
TO and clump shape, luminosity functions).
The synthetic CMDs in best agreement with the data for each source of stellar
models are shown in Fig.~\ref{simpap}, where the plotted stars are the 638
synthetic ones (30\% of which in binary systems) plus the 801 stars of the external 
field of equal area and mag limit. These models are computed without
differential reddening. The adoption of $\Delta$\ebv = 0.03 would simply widen
their TO region, leaving them in agreement with the data.
Fig.~\ref{lf} presents an example of
the observed and synthetic luminosity functions (LFs): it
compares the V band LF corresponding to the top left-hand panel CMD of 
Fig.~\ref{simpap} with that
derived from the reference CMD of Fig.~\ref{empcmd}. Similar LFs have been 
obtained for the other synthetic CMDs consistent with the data. It is apparent 
that the synthetic LF is consistent with the observational one, but not in
perfect agreement. However, the differences (at bright magnitudes) are 
completely ascribable to the observed blue stars brighter than the turn-off, 
either blue-stragglers or non members not present in our external field.

All the cases in best agreement with the data turn out to have solar 
metallicity. For the different sources of stellar tracks the best combination 
of parameters is: age 1.3 Gyr, distance modulus \mmm = 11.3 and reddening 
\ebv = 0.34 for the BBC models; age 1.3 Gyr, \mmm = 11.4 and \ebv =
 0.36 for the FST models with $\eta$=0.02; age 1.0 Gyr, \mmm = 11.3 
and \ebv = 0.38 for the FRA models. 
Bearing in mind that the former two sets of tracks take overshooting
into account while the latter does not, and that overshooting makes stellar
models brighter, the resulting ages are perfectly consistent with each other.
The FST models with larger overshooting, $\eta$=0.03, provide older
ages ($\sim$1.5 Gyr) and do not reproduce equally well clump and upper main
sequence. 

By considering all the involved uncertainties, we thus assign to NGC 6939
an age of 1.3 $\pm$ 0.1 Gyr or 1.0 $\pm$ 0.1 Gyr (depending on whether or not 
overshooting should
be assumed), a reddening of 0.34 $\leq$ \ebv $\leq$ 0.38 and a distance modulus
of 11.3 $\leq$ \mmm $\leq$ 11.4.

Models with increasingly lower metallicity predict turn-off shapes and main
sequence slopes increasingly different from the observed ones. For this reason,
even if the solar metallicity models are not perfect either, we do not consider
adequate enough those with Z$<$0.01. This result is consistent with the
literature abundances quoted in the Introduction, corresponding to
0.012$\leq$ Z $\leq$ 0.019.  

\section{Summary and discussion}

We have examined UBVI CCD data of the open cluster NGC 6939, and determined
its distance, reddening, age and (approximate) metallicity making use of
the synthetic CMDs technique and three different sets of theoretical tracks.
We found that the best combination of the cluster parameters is: 
solar abundance,
0.34 $\leq$ \ebv $\leq$ 0.38, 
11.3 $\leq$ \mmm $\leq$ 11.4, 
1.0 Gyr $\leq$ age $\leq$ 1.3 Gyr.
We also found indication of the presence of binary systems.

Our findings do not agree completely with literature data, although quite
similar in some cases (see Introduction). 
For instance, when comparing with RB02, we do not find a substantial 
differential reddening and our age is  slightly lower than theirs.
The age difference is likely due to the use of different versions of the
Padova tracks: they adopt the Girardi et al. (2000) solar metallicity
isochrones, while we prefer the older BBC tracks .  This seems to be confirmed 
by our best concordance with Carraro et al. (1998) results, based on our same
BBC tracks. Our intrinsic distance modulus is
slightly larger than that (10.95) derivable from the corrected $(m-M)_V$ = 
12.45 of Twarog et al. (1997) assuming their mean reddening \ebv = 0.47.
However, if our mean reddening \ebv = 0.36 is applied to their $(m-M)_V$, the
resulting intrinsic modulus is 11.3, identical to ours.

Regarding the differential reddening claimed by RB02, we 
cannot substantiate their values. When looking at the CMDs we derive from
our data (Sect. 3.3) we do not see any clear indication of its
presence. Furthermore, the synthetic CMDs are able to reproduce the
cluster sequences without invoking differential reddening; we think that
a $\Delta$ E(B-V) as large as
suggested by RB02 would have broadened the MS much more than observed. We
agree, however, that a small variation of $\Delta$\ebv $\leq$0.03 over
the field may allow us to well reproduce the TO morphology. 
We suggest that the presence of a fraction around 30\% of binary
stars in the cluster would further improve the CMD reproduction. 
Indeed, the broadening effect on the MS of binaries and of
differential reddening differ from each other. Binarism affects both magnitudes
and colours and leads to an increasing widening of the MS towards fainter
magnitudes, with the binary MS converging on the single MS at the TO region.
Vice versa, differential reddening shifts the colours by the same amount at all
magnitudes, thus making the effect much more evident in the brighter, tighter
TO regions.

The rather poor fit of the lower MS obtained for all the models employed is not
unheard of, and cannot be taken as a general failure to measure cluster
properties.
Very recently Grocholski \& Sarajedini (2003) presented a comparative study of
isochrone fits to CMDs in the visual and infrared filters. They took five sets
of theoretical models, compared them in the luminosity - temperature plane, and
proceeded to fit the observed CMDs of six open clusters of assorted ages and
metallicities with the isochrones transformed into the observational planes. 
The soundness of the comparison of different evolutionary codes is  probably
undermined by the fact that Grocholski \& Sarajedini (2003) use the isochrones
as transformed from the theoretical plane (luminosity and temperature) to the
observational one (magnitudes and colours) by each group, i.e. following quite
different prescriptions. This introduces additional effects,  which we 
prefer to avoid using a single transformation for all tracks.
Unfortunately, we have no theoretical set in common with them (they use yet
another version of the Padova tracks, by Girardi et al. 2002), but their
results are quite interesting: there is no single set that is able to match
all CMDs in all colours, at all ages, and for all metallicities. 
All of them behave rather well in some cases, and do not fit well in others.
However, when the observed CMDs are deep enough, the fit to the lower main
sequence is often (but curiously not always) bad; this happens in general
around M$_{\rm V} \simeq $ 8 (the same we find for NGC 6939) for the visual
colours, and is attributed by 
Grocholski \& Sarajedini (2003) to some still missing ingredient in the
model atmospheres for low mass stars.

Finally, we wish to stress that the roughly solar abundance we find  for NGC
6939 is more an indication than a well founded measurement. It is encouraging
that it's in  reasonable agreement with what other methods find, but high
resolution spectroscopy and detailed abundance analysis should be performed to
determine the cluster elemental composition.


\bigskip\noindent
ACKNOWLEDGEMENTS

This work is based on observations made with the Italian
Telescopio Nazionale Galileo (TNG) operated on the island of La Palma by the
Centro Galileo Galilei of the INAF (Istituto Nazionale di Astrofisica) at the
Spanish Observatorio del Roque de los Muchachos of the Instituto de Astrofisica
de Canarias. We thank M. Bellazzini for the data acquired on August 2001 and
P. Ventura for the FST models.
We are grateful to B. Twarog for having informed us of the correct value of
their distance modulus and for having pointed out a problem of ours.
The bulk of the simulation code was originally provided by L.Greggio.
Financial support to this project has come from the MURST-MIUR through 
Cofin98 ''Stellar Evolution'', and Cofin00 ''Stellar Observables of Cosmological
Relevance''.
This research has made use of the Simbad database, operated at CDS, Strasbourg,
France. 
Finally we acknowledge the use of the valuable BDA database, maintained by
J.-C. Mermilliod, Geneva.

\end{document}